\documentclass{PoS}
\newcommand{\be}{ \begin{eqnarray}}
\newcommand{\ee}{\end{eqnarray} }
\newcommand{\Tr}{{\rm Tr}}

\def\MSbar{{\rm  \overline{\footnotesize MS\kern-0.05em}\kern0.05em}}
\newcommand{\Dlr}{\buildrel \leftrightarrow \over D\raise-1pt\hbox{}}
 \newcommand{\Dl}{\buildrel \leftarrow \over D\raise-1pt\hbox{}}
\newcommand{\Dr}{\buildrel \rightarrow \over D\raise-1pt\hbox{}}

\newcommand{\pslash}{p \!\!\!/}

\newcommand{\msbar}{\overline{\mbox{MS}}}
\newcommand{\Ort}{{\rm O\,}}
\newcommand{\Hy}{{\rm H\,}}
\def\VEV#1{\langle #1 \rangle}

\title{Renormalization constants for $N_{\rm f}=2+1+1$ twisted mass QCD}
 
\ShortTitle{Renormalization constants for $N_{\rm f}=2+1+1$ twisted mass QCD}

\author{Benoit Blossier\\
   Laboratoire de Physique Th\'eorique - CNRS et Universit\'e  Paris-Sud XI \\
B\^atiment 210, 91405 Orsay Cedex France
     }
  \vspace*{-0.3cm}    
 \author{Mariane Brinet\\
   Laboratoire de Physique Subatomique et de Cosmologie - CNRS/IN2P3/UJF/INPG, 
53, avenue des Martyrs, 38026 Grenoble, France
    }       
  \vspace*{-0.3cm}    
\author{Pierre Guichon\\
CEA-Saclay, IRFU/SPhN, 91191 Gif-sur-Yvette France\\
    }
  \vspace*{-0.5cm}    
\author{Vincent Mor\'enas\\
       Laboratoire de Physique Corpusculaire, Universit\'e Blaise Pascal, CNRS/IN2P3 
63177 Aubi\`ere Cedex, France\\
     }
  \vspace*{-0.5cm}    
\author{Olivier P\`ene\\
   Laboratoire de Physique Th\'eorique - CNRS et Universit\'e  Paris-Sud XI \\
B\^atiment 210, 91405 Orsay Cedex France
      }
    \vspace*{-0.3cm}          
        \author{Jose Rodr\'iguez-Quintero\\
 Dpto. F\'isica Aplicada - Fac. Ciencias Experimentales,\\
Universidad de Huelva 21071 Huelva Spain
       }
   \vspace*{-0.3cm}           
\author{\speaker{Savvas Zafeiropoulos}\\
       Laboratoire de Physique Corpusculaire, Universit\'e Blaise Pascal, CNRS/IN2P3 
63177 Aubi\`ere Cedex, France and\\
Institut f\"ur Theoretische Physik, Goethe-Universit\"at Frankfurt,\\
Max-von-Laue-Str.~1, 60438 Frankfurt am Main, Germany\\
        E-mail: \email{zafeiro@th.physik.uni-frankfurt.de}}

\abstract{We summarize recent non-perturbative results obtained for the renormalization constants computed in the RI'-MOM scheme for $N_{\rm f}=2+1+1$ twisted mass QCD. Our implementation employs the Iwasaki gauge action and four dynamical degenerate twisted mass fermions. Renormalization constants for scalar, pseudo-scalar, vector and axial operators, as well as the quark propagator renormalization, are computed at three different values of the lattice spacing, two different volumes and several values of the twisted mass. Our method  allows for a precise cross-check of the running, because of the particular proper treatment of the hypercubic artifacts. Preliminary results for twist-2 operators are also presented.}

\FullConference{The 32nd International Symposium on Lattice Field Theory\\
                 23-28 June, 2014\\
                 Columbia University New York, NY}
\let\ifpdf\relax
\begin{document}

\section{Introduction}
\vspace*{-0.25cm}
Quantum Chromodynamics (QCD) is strongly coupled in the scales relevant for Hadron Physics and thus requires a non-perturbative treatment. At the moment the only approach which avoids approximations and modelling assumptions is QCD discretized on a spacetime lattice. However, lattice QCD is a bare field theory where the results of matrix elements are computed at a fixed value of the UV cutoff $\Lambda=1/a$. One must renormalize in order to obtain continuum physics and this can be done perturbatively or non-perturbatively. Lattice perturbation theory is infamous for its slow convergence properties so we will employ the non-perturbative (NP) method that fits naturally to the whole non-perturbative setup of the computation. Of course the lattice computation is hindered by $\mathcal{O}(a^2)$ discretization effects if e.g. one employs a Symanzik improved action or the Twisted mass (TM) formulation. The two main players in the game of NP renormalization are the RI'-MOM scheme \cite{Martinelli} and the Schr\"odinger functional \cite{SchFunctional}.
There has been a lot of work utilizing the RI'-MOM scheme \cite{Gockeler} and a significant amount of it has been within the framework of TM QCD \cite{CyprusLocal,CyprusTwist2,Zq, Palao,Italiani,Francesco}. 
In this study we employ the RI'-MOM scheme to compute the renormalization constants (RCs) of fermionic bilinears for $N_{\rm f}=4$ 
using TM fermions. We perform the renormalization of the densities, the currents as well as of the twist-2 operator $\displaystyle O_{44}(x)\propto\bar{u}(x)[\gamma_4\Dlr_4-\frac{1}{3}\sum_k\gamma_k\Dlr_k]u(x)$ which determines the average momentum fraction $\langle x\rangle_q =\int_0^1 dx \;x \left(q (x) +\bar q(x)\right)$ in the hadrons \cite{Negele}.  
Our methods allow for the extraction of the $\langle A^2\rangle$, the dimension-2 gluon condensate that has rich phenomenological implications.

\section{RI'-MOM and the computational setup}
\vspace*{-0.25cm}
We use the RI'-MOM scheme \cite{Martinelli}
and we focus on local fermion bilinears \be O_{\Gamma}=\bar{\psi}(x_1)\Gamma\psi(x_2),\ee 
where $\Gamma$ can be any Dirac structure and contains covariant derivatives for the case of twist-2 operators.
We insert  $O_{\Gamma}$ in the fermion 2-pt function, which determines the 3-pt function \be G_O=\langle u(x_1)O_{\Gamma}\bar{d}(x_2)\rangle ,\ee
then we compute the amputated Green's function, or bare vertex, 
as \be\Lambda_O(p)=S^{-1}_u(p)G_O(p)S^{-1}_d(p) ,\ee with $S(p)$ the quark propagator.
The operators that we consider are multiplicatively renormalized.
Since the RC is a scalar, one works with the projected quantity 
\be \Gamma_O(p)=\frac{1}{12}\Tr[P_O\Lambda_O(p)],\ee
where $P_O$ is a projector specific for each operator.\\ The operator RC ($Z_O$) is fixed by the following renormalization condition

\be 
\Gamma_O (\mu,g_R,m_R=0)  \big |_R = 
 Z_q^{-1}(a\mu,g_0)  Z_O(a\mu,g_0)  \Gamma_O (p,g_0,m) 
 \big |_{\begin{array}{cc} p^2 = \mu^2 \\ m \to 0 \end{array}}  =  1,
\ee
after having computed the quark wave function RC which is given
\vspace*{-0.65cm} by \\
\be  Z_q(\mu^2=p^2)=-\frac{i}{12p^2}\Tr[S^{-1}_{bare}(p)\pslash]. \ee
\vspace*{-0.05cm}
Unity on the RHS of the renormalization condition represents the tree level value of $\Gamma_0$.
Note that since the 2-pt and 3-pt correlation functions are not gauge invariant, one needs to fix the gauge for the 
determination of the RCs. We have chosen the lattice version of the Landau gauge in this study.
In order to obtain sensible results through this renormalization procedure one needs to be within the window of applicability of the RI'-MOM scheme which is defined in the case of a lattice regularization as follows
\vspace*{-0.5cm}
\be\Lambda_{QCD}\ll \mu\ll \frac{\pi}{a},\ee
where the first inequality ensures the possibility of matching with some perturbative scheme like the $\msbar$ and protects from infrared effects such as Goldstone pole contaminations 
while the second inequality ensures small cutoff effects. With the available resources the statistical error in the computation of RCs is rather miniscule while the systematic errors are originating mainly from cutoff effects which are considerably larger. Thus we have tried to isolate the different cutoff effects which contaminate our results and we treat them non-perturbatively utilizing group theoretical methods \cite{Feli, H4}.       
There are two dominant type of cutoff effects. Those which are invariant under $\Ort(4)$ and those that are only invariant under $\Hy (4)$, the group of hypercubic rotations in four dimensions.

In the current computations of $N_{\rm f}=2+1+1$ flavors the ETMC is employing the Iwasaki action in the gauge sector and a twisted mass  action for the heavy as well as the light quarks \cite{Frezzossi}. 
The full action reads
 \be S=S^{\tiny {YM}}_{Iwa}+S^f_l+S^f_h,\ee
while the fermionic part of the action  \cite{Baron}
\be S^f_l+S^f_h&=&a^4\displaystyle\sum_{x}\bar{\chi}_l\left(\gamma \cdot\nabla-\frac{a}{2}\nabla\cdot\nabla+m_{0l}+i\mu_l\gamma_5\tau_3\right)\chi_l\nonumber\\ &+&a^4\displaystyle\sum_{x}\bar{\chi}_h\left(\gamma \cdot\nabla-\frac{a}{2}\nabla\cdot\nabla+m_{0h}+i\mu_h\gamma_5\tau_1+\mu_{\delta}\tau_3\right)\chi_h.
\ee
The polar mass is defined as $M=\sqrt{m^2+\mu^2}$ and the twist angle as $\omega=\arctan(\mu/m)\;$ where $m=Z_Am_{\rm PCAC}$. 
The quark doublet in the twisted basis is related to the one in the physical basis by the transformation
\vspace*{-0.45cm}
\be\psi_l&=&e^{{i\over 2}\omega_l\gamma_5\tau_3} \chi_l\nonumber,\\ \bar{\psi}_l&=&\bar{ \chi}_le^{{i\over 2}\omega_l\gamma_5\tau_3}\nonumber,\\
\psi_h&=& e^{-i \omega_1 \gamma_5 \tau_1/2} e^{i \omega_2 \tau_2/2} \chi_h,\nonumber\\
\bar{\psi}_h&=&\bar{\chi}_h e^{- i \omega_2 \tau_2/2}e^{-i \omega_1 \gamma_5 \tau_1/2}.\ee
which transforms the action to the conventional Dirac action
\be S_{ph}=a^4\displaystyle\sum_{f=h,l}\bar{\psi}_f\left(D_{\rm tW} +M_f\right)\psi_f.\ee
In order to achieve the benefits of the TM formulation, such as automatic $\mathcal{O}(a)$ improvement, one needs to work at maximal twist $\omega=\pi/2$ 
\cite{Frezzossi} which amounts to tuning $m_{PCAC}$ to zero.
For the $N_{\rm f}=4$ configurations this tuning was a highly non trivial task at the time these configurations where produced.
So an alternative strategy was followed, to average results obtained at two opposite values of $m_{PCAC}$.
The combination of the results with positive and negative $m_{PCAC}$ that goes usually under the name of $\theta$ average is the first step of our calculations in order to get rid of the $\mathcal{O}(a)$ cutoff effects.
The $N_{\rm f}=4$ configurations were generated with the main purpose to renormalize accurately the physical $N_{\rm f}=2+1+1$
configurations. The reason is that since we employ a mass independent renormalization scheme (where RCs are defined in the chiral limit), the $N_{\rm f}=4$ ensembles with four light degenerate flavors allow for a reliable chiral extrapolation.
In this analysis the configurations used comprise two different volumes, three values of the lattice spacing, as well as several values of the twisted mass.
The values of the lattice spacing are respectively $a=0.062$ fm for $\beta=2.10$, $a=0.078$ fm for $\beta=1.95$  and $a=0.086$ fm for $\beta=1.90$ \cite{Palao}.
 The exact parameters of the runs are 
summarized in Table \ref{tab1}.  
\vspace*{-0.25cm}
\begin{table}[htdp]
\caption{{\it{ The $N_f=4$ ETMC ensembles utilized in this study note that $a\mu_{sea}$ is given in bold}}}
\vspace*{-0.35cm}
\begin{small}
\begin{center}
\begin{tabular}{|l|c|c|c|c|}\hline
ensemble  & $\kappa$  & $am_{PCAC}$& $a\mu$   & confs $\#$ \\\hline
\hline
\multicolumn{5}{|c|}{$\beta=2.10$ - $32^3.64$}\tabularnewline
\hline
3p &  0.156017 &+0.00559(14) &0.0025, {\bf{0.0046}}, 0.0090, 0.0152, 0.0201, 0.0249, 0.0297  & 250   \\\hline
3m &  0.156209 &-0.00585(08) &0.0025, {\bf{0.0046}}, 0.0090, 0.0152, 0.0201, 0.0249, 0.0297  & 250   \\\hline
4p &  0.155983 & +0.00685(12)&0.0039, {\bf{0.0064}}, 0.0112, 0.0184, 0.0240, 0.0295  & 210   \\\hline
4m &  0.156250 & -0.00682(13)&0.0039, {\bf{0.0064}}, 0.0112, 0.0184, 0.0240, 0.0295  & 210   \\\hline
5p &  0.155949 &+0.00823(08)  &0.0048,  {\bf{0.0078}}, 0.0119, 0.0190, 0.0242, 0.0293 & 220   \\\hline
5m &  0.156291 &-0.00821(11) & 0.0048,  {\bf{0.0078}}, 0.0119, 0.0190, 0.0242, 0.0293 & 220   \\\hline
\hline
\multicolumn{5}{|c|}{$\beta=1.95$ - $24^3.48$}\tabularnewline
\hline
2p &  0.160826 &+0.01906(24) &{\bf{0.0085}}, 0.0150, 0.0203, 0.0252, 0.0298  & 290   \\\hline
2m &  0.161229 &-0.02091(16)  &{\bf{0.0085}}, 0.0150, 0.0203, 0.0252, 0.0298  & 290   \\\hline
3p &  0.160826 & +0.01632(21)&0.0060, 0.0085, 0.0120, 0.0150, {\bf{0.0180}}, 0.0203, 0.0252, 0.0298  & 310   \\\hline
3m &  0.161229 & -0.01602(20)&0.0060, 0.0085, 0.0120, 0.0150, {\bf{0.0180}}, 0.0203, 0.0252, 0.0298  & 310   \\\hline
8p &  0.160524 &+0.03634(14) &{\bf{0.0020}}, 0.0085, 0.0150, 0.0203, 0.0252, 0.0298& 310   \\\hline
8m &  0.161585 &-0.03627(11)   & {\bf{0.0020}}, 0.0085, 0.0150, 0.0203, 0.0252, 0.0298 & 310   \\\hline
\hline
\multicolumn{5}{|c|}{$\beta=1.90$ - $24^3.48$}\tabularnewline
\hline
1p &  0.162876 &+0.0275(04) &0.0060, {\bf{0.0080}}, 0.0120, 0.0170, 0.0210, 0.0260  &   450  \\\hline
1m&  0.163206 &-0.0273(02) &0.0060, {\bf{0.0080}}, 0.0120, 0.0170, 0.0210, 0.0260  &   450  \\\hline
4p &  0.162689 &+0.0398(01) &0.0060, {\bf{0.0080}}, 0.0120, 0.0170, 0.0210, 0.0260  &   370  \\\hline
4m&  0.163476 &-0.0390(01) &0.0060, {\bf{0.0080}}, 0.0120, 0.0170, 0.0210, 0.0260  &   370  \\\hline
\end{tabular}
\end{center}
\end{small}
\label{tab1}
\end{table}%
\vspace*{-0.5cm}
The next step in the analysis is to take the chiral limit in the valence sector where we also take care of the Goldstone pole in the case of the pseudoscalar RC. 
We use the following ansatz for the amputated pseudoscalar vertex
\be\Gamma_P=a_P+b_Pm^2_{\pi}+\frac{c_P}{m_{\pi}^2},\ee
and we "subtract" the pole contributions according to 
\be\Gamma_P^{sub}=\Gamma_P-\frac{c_P}{m_{\pi}^2}.\ee
\vspace*{-0.35cm}
\section{Correcting for discretization effects}
\vspace*{-0.15cm}
With the improvement provided by the TM formulation we still have $\mathcal{O}(a^2)$ lattice artifacts contaminating our results. We will correct for the $\Hy (4)$, hypercubic artifacts by employing the "$H(4)$"-extrapolation method of \cite{Feli} which treats them non-perturbatively.
From basic group theory \cite{Weyl} one can show that any $\rm H\,(4)$ invariant polynomial can be expanded in the basis of the following $\rm H\,(4)$ invariants
\be p^{[2]}=\displaystyle\sum_{\mu=1}^{4}p_{\mu}^2,\;\;\; p^{[4]}=\displaystyle\sum_{\mu=1}^{4}p_{\mu}^4,\;\;\; 
 p^{[6]}=\displaystyle \sum_{\mu=1}^{4}p_{\mu}^6,\;\;\;  p^{[8]}=\displaystyle \sum_{\mu=1}^{4}p_{\mu}^8...\ee    
  
We start by expanding the RC already averaged over the cubic orbits around $p^{[4]}=0$ as,
\be Z_{latt} (a^2 p^2, a^4p^{[4]}, a^6p^{[6]}, ap_4, a^2\Lambda^2_{QCD})=Z_{hyp_{}corrected} (a^2 p^2, ap_4, a^2\Lambda^2_{QCD})+
R(a^2p^2,a^2\Lambda_{QCD}^2)\frac{a^2p^{[4]}}{p^2}+ \ldots \nonumber\\
&&
\ee
\vspace*{-0.5cm}
where
\vspace*{-0.5cm}
\be R(a^2p^2,a^2\Lambda^2_{QCD})=\frac{dZ_{latt} (a^2 p^2, 0, 0, 0, a^2\Lambda^2_{QCD})}{d\epsilon}|_{\epsilon=p^{[4]}/p^2\ll 1}=c_{a2p4}+c_{a4p4}a^2p^2.\ee
The effect of the $H(4)$ corrections can be seen in Figs.[1,2].

\vspace*{-0.15cm}
\begin{figure}
\vspace*{0.35cm}
\begin{minipage}{.5\textwidth}
  \centering
  \includegraphics[width=.95\linewidth]{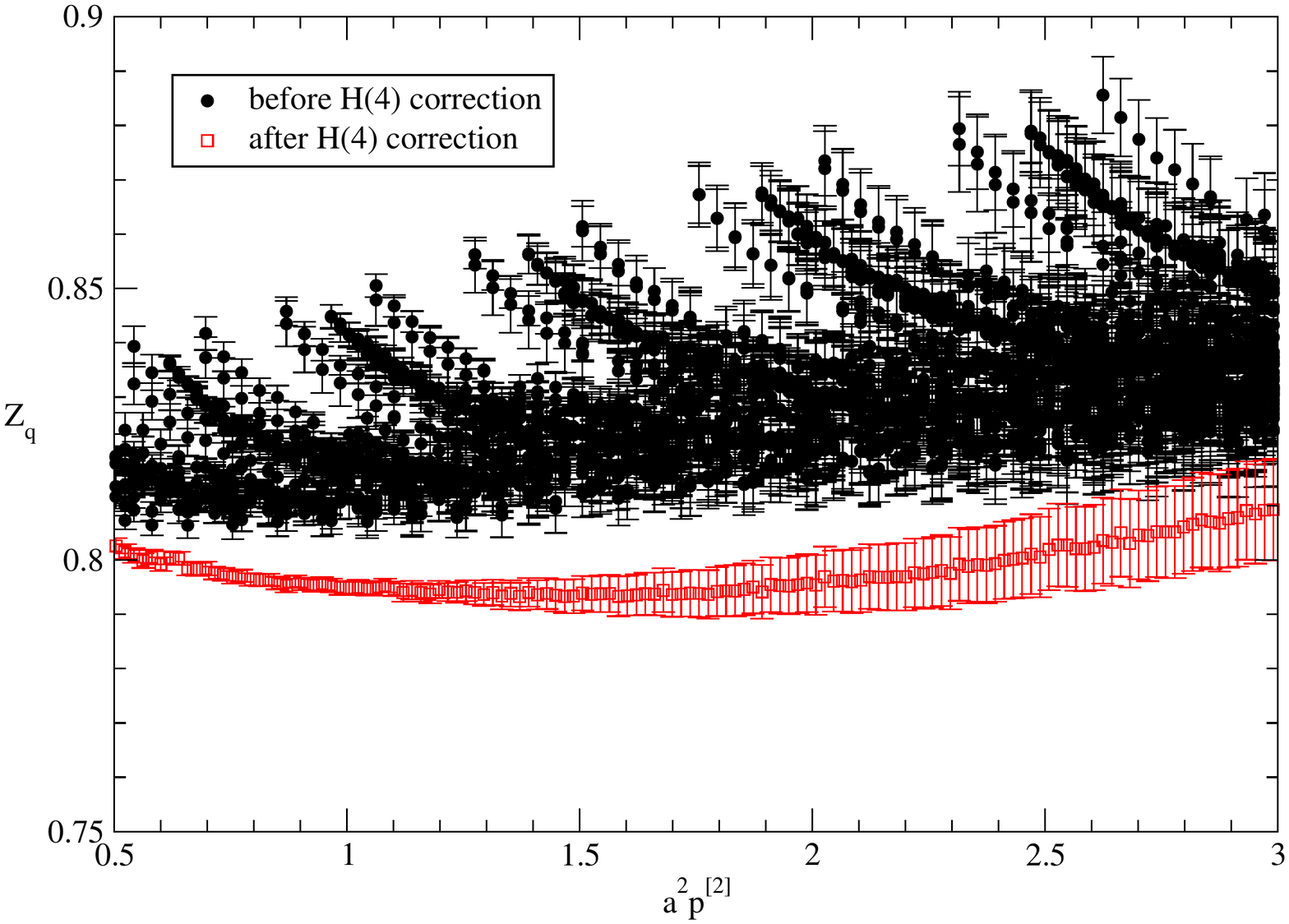}
  \vspace*{-0.85cm}
  \caption{\label{fig1} The half fishbone structure arises because of the hypercubic artifacts. The same value of $p^2$ corresponds to different hypercubic orbits (black data). The "$H(4)$ -  extrapolation" (red data) treats these cutoff effects non perturbatively. Results are shown for $\beta=2.10$, volume $32^3\times64$ and $\mu=0.0046$.}
\end{minipage}%
\hspace*{0.35cm}
\begin{minipage}{.5\textwidth}
  \centering
  \includegraphics[width=.99\linewidth]{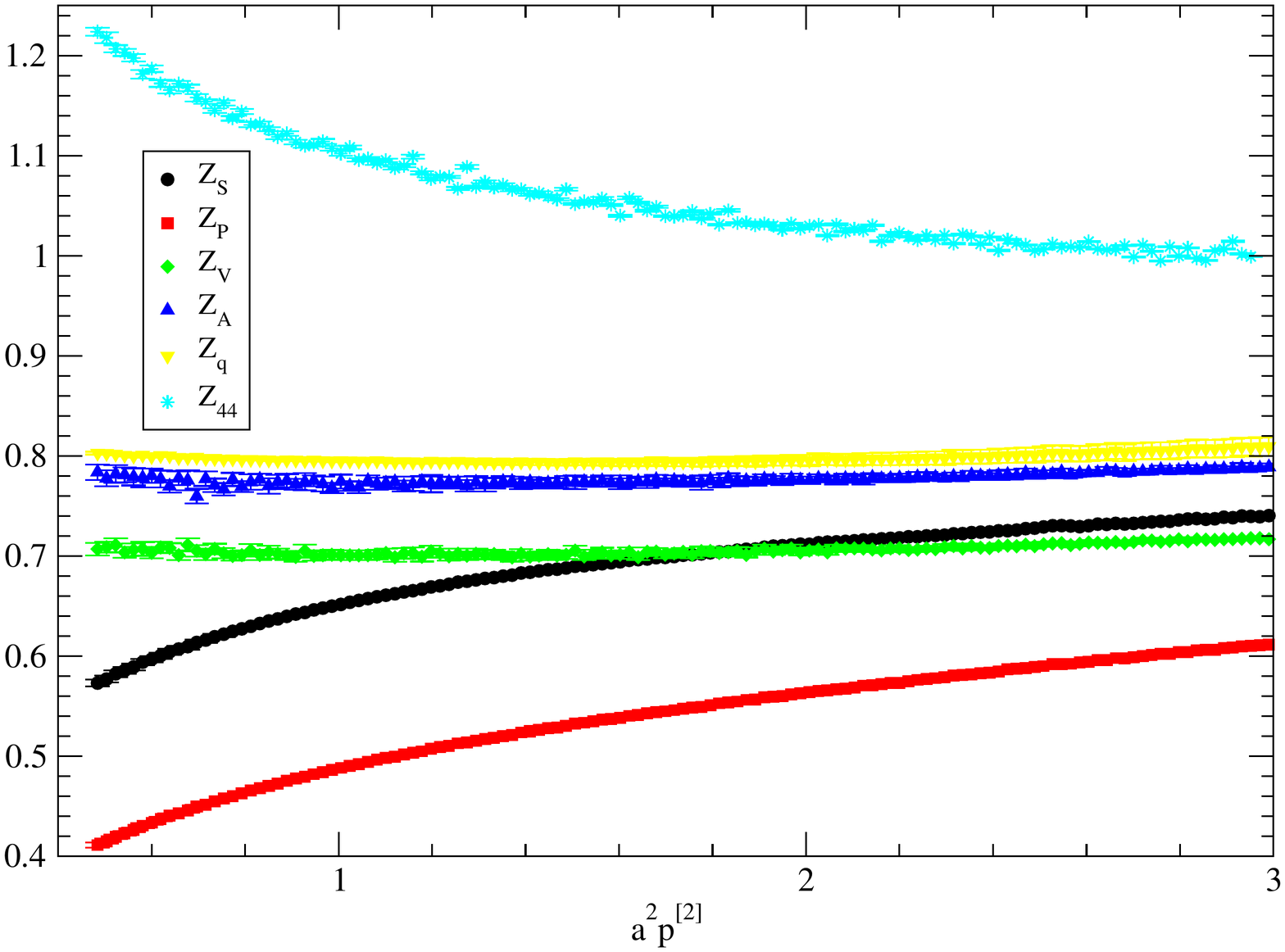}
  \caption{\label{fig2}RCs as a function of $a^2p^2$ after the removal of the $H(4)$ artifacts for the ensemble with $\beta=2.10$, volume $32^3\times64$ and $\mu=0.0046$.}
  \label{fig:test2}
\end{minipage}
\end{figure}
Then in order to take into account the lattice artifacts which are invariant under $O(4)$
we explicitly add $O(4)$ lattice artifacts in the running of the RCs. In the case of the quark wave function renormalization we utilize the OPE inspired formula
 for the perturbative running of $Z_q$ \cite{Zq}
 \begin{eqnarray}\label{Eq:Zqrun1}
Z_q^{hyp-corr}(a^2p^2)&=& Z^{pert\,RI'}_q(\mu^{2})\,c^{RI'}_{0Z_q}({p^2\over \mu^{2}},\alpha(\mu))\nonumber \\
&\times&\left( 1 +   \frac{ \VEV{A^2}_{\mu^2}}{32p^2}
\frac{c_{2Z_q}^{\msbar}({p^2\over \mu^2},\alpha(\mu))}
{c_{0Z_q}^{RI'}({p^2\over \mu^2},\alpha(\mu))}\frac{c_{2Z_q}^{RI'}({p^2\over \mu^2},\alpha(\mu))}
{c_{2Z_q}^{\msbar}({p^2\over \mu^2},\alpha(\mu))} \right)\nonumber \\
&+&  c_{a2p2}\; a^2\,p^2 +  c_{a4p4}\;(a^2p^2)^2 .
\end{eqnarray} 
The coefficients $c_{0Z_q}^{RI'}$, $c_{0Z_q}^{RI'}$ and $c_{2Z_q}^{\msbar}$  are known perturbatively \cite{ ChetyrkinRetey, 3loops}. 
While $Z^{pert\,RI'}_q(\mu^{2})$ ,  $\VEV{A^2}_{\mu^2}$,  $c_{a2p2}$ and $c_{a4p4}$
will be determined through fitting, see Fig.~3. Note the presence of the gluon condensate, $\VEV{A^2}_{\mu^2}$, which plays a crucial role in the RC of gauge variant quantities such as the quark field and since also the whole computation takes place in a gauge fixed setting.
\begin{figure}[t]
\center
 \includegraphics[height=4cm]{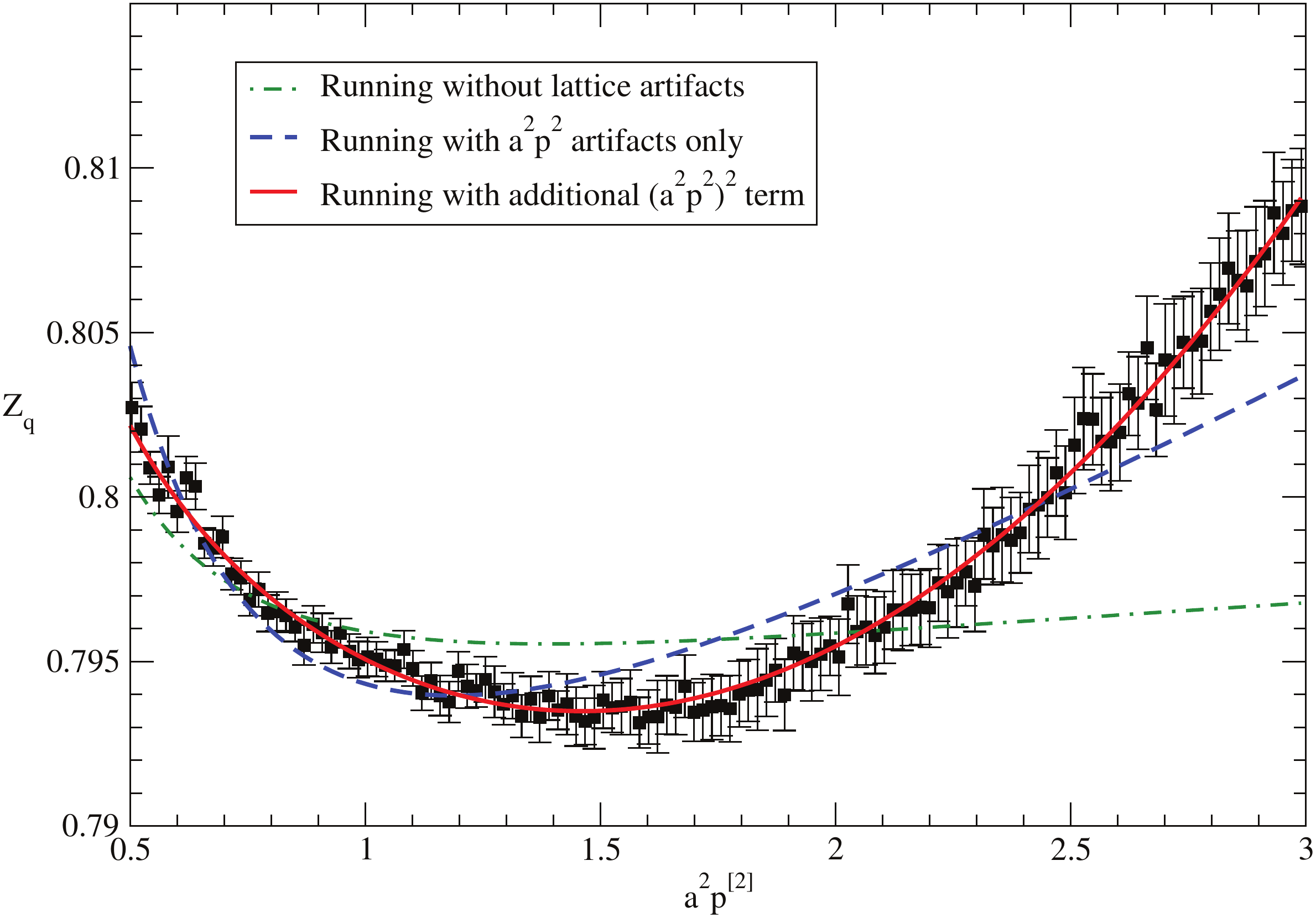}
 \caption{\label{fig3} The running of $Z_q$ for the ensemble 3mp.}
\end{figure}
One can immediately see that the running formula contains $O(4)$ invariant lattice artifact terms $\propto a^2p^2$ and $\propto (a^2p^2)^2$.
We use similar running formulae for all the RCs and we refer the reader to \cite{ourNf4} for all the details. Note that for the other composite operators we might have terms $\propto 1/(a2p2)$ which also mainly accounts for the gluon condensate for $Z_P, Z_S\;$ and $Z_{44}$. The perturbative running allows us to run the RCs up to large scales, such as, $10 {\rm GeV}$ and at this point we use perturbative formulae \cite{Gracey} to convert our results to the $\msbar$ scheme which will allow us to make contact with phenomenological calculations and experiments. In the following table we tabulate all the results for the RCs in the $\msbar$ scheme at $2 {\rm GeV}$.
\vspace*{0.24cm}
\hspace*{-3.4cm}
\begin{table}[htbp]
\hspace*{-2.7cm}
\begin{tabular}{c|ccccccc}\hline\hline
$\beta$    & $Z_q$                      &    $Z_S$                              &  $Z_P$               &     $Z_V$              &        $Z_A$                            & $Z_P/Z_S$    & $Z_{44}$ \\\hline\hline
1.90         & 0.762(3)(5)(2)         &  0.722(3)(5)(9)          &     0.431(3)(3)(6)          & 0.623(2)(1)(5)               &0.717(1)(2)(4)                     & 0.597(4)(4)    (3)&  0.973(9)(7)(30)    \\
1.95         & 0.770(2)(6)(6)     & 0.722(4)(5)(3)              &0.461(2)(4)(5)                &0.639(2)(1)(4)    &      0.726(2)(2)(4)                &0.638(4)(4)(3)    &  0.977(12)(11)(30)  \\
2.10         & 0.787(2)(6)(6)     & 0.725(2)(5)(3)              & 0.522(1)(4)(1)               &0.687(1)    (1)(2)     &    0.755(1)(2)(4)                   &0.720(4)(2)(5)     & 1.019(8)(6)(30) \\\hline\hline
\end{tabular}
\caption{Final results for $N_f=4$ RCs in the $\msbar$ scheme at 2 GeV. We have quoted the statistical error in the first parenthesis, while in the second parenthesis the systematic error due to cutoff effects and in the third parenthesis we quote the systematic error originating from the chiral extrapolation.\label{Table:msbar_syst}}
\end{table}  

\vspace*{-0.75cm}
\section{Conclusions and Outlook}
\vspace*{-0.25cm}
We have presented our results for the RCs of the quark propagator, densities and currents and for the $O_{44}$ twist-2 operator for $N_f=4$ twisted mass fermions. 
We have implemented a systematic and rigorous procedure for the correction of the hypercubic lattice artifacts. 
We have treated the main source of uncertainty, the cutoff effects in a non perturbative way for the  $O(4)$ breaking artifacts as well with an OPE inspired perturbative formula for the $O(4)$ invariant artefacts.
For a comparison with experiments and phenomenological calculations our results, obtained in the RI'-MOM scheme, 
have been converted to the  $\msbar$ scheme at $2 {\rm GeV}$. 
We plan to apply our method to other twist-2 operators containing more than one covariant derivatives as well as to the new configurations that will be produced shortly by the ETMC at almost the physical point.\\
\vspace*{0.3cm}
{\bf Acknowledgments.}\\
\vspace*{-0.05cm}
We would like to thank P. Boucaud, N. Carrasco, P. Dimopoulos, R. Frezzotti, V. Lubicz, S. Simula and T. Vladikas for fruitful discussions.
This work was granted access to the HPC resources of CINES and IDRIS under the allocations 2013-052271 and 2014-052271 made by GENCI. Propagator computations have also extensively used CINECA GPUs in the framework of the DECI-9 project DEC09-NPR-LQCD. Most of the analysis took place
in Lyon-CCIN2P3. We express our gratitude to the staff of all these Computing facilies for their constant help. 
This work was supported by the CNRS and the Humboldt Foundation (S.Z.).
\vspace*{-0.25cm}


\end{document}